\documentclass[global,twocolumn]{svjour}
\usepackage{graphicx}
\journalname{Applied Physics B}
\begin{document}
\title{Mixture of ultracold lithium and cesium atoms in an optical dipole trap}
\author{A. Mosk\and S. Kraft\and M.
Mudrich\and K. Singer \and W. Wohlleben\thanks{\emph{Present
address:} Max-Planck-Institut f\"ur Quantenoptik, 85748 Garching,
Germany }\and R. Grimm\thanks{\emph{Permanent address:} Institut
f\"{u}r Experimentalphysik, Universit\"at Innsbruck, A-6020
Innsbruck, Austria}\and and M. Weidem\"{u}ller
}                     
%
%
\institute{Max-Planck-Institut f\"{u}r Kernphysik, 69029 Heidelberg,  Germany
 }
\date{Received: date / Revised version: date}
%
\maketitle
\begin{abstract}
We present the first simultaneous trapping of two different
ultracold atomic species in a conservative trap. Lithium and
cesium atoms are stored in an optical dipole trap formed by the
focus of a CO$_2$ laser. Techniques for loading both species of
atoms are discussed and observations of elastic and inelastic
collisions between the two species are presented. A model for
sympathetic cooling of two species with strongly different mass in
the presence of slow evaporation is developed. From the observed
Cs-induced evaporation of Li atoms we estimate a cross section for
cold elastic Li-Cs collisions.
\end{abstract}
\section{Introduction}
\label{intro}

Mixtures of ultracold atoms add many new possibilities to the
study of interactions in ultracold gases since the different
components of the mixture can have widely different intrinsic
properties, e.g., different masses, spins, scattering properties
and quantum statistics. On one hand, many interesting phenomena
are exclusive to gas mixtures, such as mixing and de-mixing
phenomena~\cite{ketterle98,bigelow98}, the subtle interplay
between bosonic and fermionic atoms~\cite{molmer98}, or the
prospect of forming ultracold heteronuclear molecules through
photoassociation~\cite{wang98}. On the other hand, one may see an
atomic gas mixture as a tool for transferring the favorable
thermodynamic properties of one component (low temperature, high
magnetization) to the other component. In this case one would like
to use one gas that is easy to cool or polarize, and bring it in
contact with a second gas that is harder to handle. These
processes are commonly referred to as sympathetic cooling and
polarization exchange.

Already 15 years ago, cryogenically cooled helium was used as a
buffer gas to precool and load atomic hydrogen into a magnetic
trap~\cite{hess87}. More recently, this method was extended to
other atomic species and even molecules~\cite{doyle98}.
Sympathetic cooling with ultracold, trapped gases has initially
been studied for different \textit{hyperfine} states of
evaporatively cooled rubidium atoms in order to create two
overlapping Bose-Einstein Condensates (BECs)~\cite{Wieman97}.
Later, evaporative cooling of a mixture of \textit{magnetic spin}
states was employed as the key step to create a degenerate Fermi
gas of fermionic $^{40}$K ~\cite{DeMarco99}. Most recently,
sympathetic cooling  in magnetic traps has been applied to
different \textit{isotopes} of the same atomic species: the
bosonic species $^{85}$Rb cooled by $^{87}$Rb \cite{Esslinger} and
the fermionic $^{6}$Li cooled by $^{7}$Li \cite{ENSLi,Truscott}.
So far, no experiments are reported which investigate elastic
interactions in two distinct atomic \textit{species}.

Sympathetic cooling of a mixture of different atomic species in
principle offers an enormous advantage if one species can be
optically cooled to much lower temperatures than the other. We are
investigating a mixture of the alkaline atoms lithium and cesium.
This combination represents the prototype of a mixture in which
one component (lithium) is hard to cool optically while the other
(cesium) allows sub-Doppler temperatures ( $< 3\,\mu$K in free
space) to be reached by simple polarization-gradient cooling. In
contrast to evaporative cooling, the cooling process does not lead
to a significant loss of either the cooled gas or the cooling
agent. Sympathetic cooling with cesium (or other sub-Doppler
cooled atoms) may be the method of choice for cooling many other
atomic and molecular species, including LiCs molecules formed by
photoassociation from the same mixed-species sample!

We present the first observation of simultaneous trapping of two
different atomic species in a conservative trap: lithium and
cesium atoms stored in a very far-detuned optical dipole trap. Our
optical trap is based on a CO$_2$ laser emitting light at 10.6
$\mu$m~\cite{takekoshi96}. The trap provides a conservative
potential which is independent of the hyperfine state and spin
projection of the atoms~\cite{grimm00}. This allows us to study
exoergic spin exchange collisions as well as elastic interactions.
Exoergic collisions between different species have been previously
investigated in light-pressure
traps~\cite{Bagnato,Bigelow,schloeder99} but the near-resonant
light is known to influence the cross-sections even for collisions
involving only the ground state~\cite{gensemer00}. No previous
measurements of two-species collision cross-sections in
conservative traps exist. If all atoms are in the lower hyperfine
state, exoergic collisions are impossible. We nevertheless see
slow evaporation of lithium, which is an indication that
thermalization through elastic collisions is taking place. To
interpret this evaporation and to estimate a collison cross
section from it, we adapt a model given in
Ref.~\cite{AspectPRA2001} and we develop a rate equation model for
sympathetic cooling with slow evaporation.

This paper is organized as follows: An overview of the apparatus
is given in section~\ref{sect:quest} and the methods for loading and
characterization of the trap are discussed in section~\ref{sect:storage}.
 The exoergic spin-changing collisions
are studied in section~\ref{sect:spinchange}. Finally, a rate
equation model for sympathetic cooling is given in section~\ref{sect:symp},
and compared to numerical results and experimental data.

\section{Quasi-electrostatic trap}
\label{sect:quest}

The focus of a CO$_2$ laser beam constitutes an almost perfect
realization of a conservative trapping potential for atoms. The
atoms experience an optical dipole force which points towards the
maximum of the intensity thus providing stable confinement in all
spatial directions. Since the frequency of the CO$_2$ laser is far
below any dipole-allowed atomic transition frequency, the induced
atomic dipole moment oscillates in phase with the oscillating
electric field of the laser as if the field was a slowly varying
static field. Therefore, the name quasi-electrostatic trap (QUEST)
has been coined for this particular realization of an optical
dipole trap. Due to the large detuning of the laser frequency from
any atomic resonances, heating through photon scattering can be
completely neglected.

For an intensity distribution $I(\mathbf{r})$ the potential
$U(\mathbf{r})$ is given by $U(\mathbf{r}) = -\alpha_\mathrm{stat}
I(\mathbf{r}) / 2 \varepsilon_0 c$ where $\alpha_\mathrm{stat}$
denotes the static polarizability of the atoms
\cite{takekoshi96,grimm00}. For a focused beam of power $P$ and
waist $w$ one gets a trap depth $\epsilon_t = 2 \alpha_{\rm stat}
P/(\pi \varepsilon_0 c w^2)$. We employ an industrial, sealed-tube
CO$_2$ laser (Synrad Evolution) providing 140\,W of power in a
nearly TEM$_{00}$ transversal mode characterized by $M^2 = 1.2$.
The laser beam is first expanded by a telescope, and then focused
into the vacuum chamber by a lens of 254\,mm focal length. The
focus has a waist of 85\,$\mu$m, yielding a trap depth of
390\,$\mu$K for Li and 1000\,$\mu$K for Cs (in units of the
Boltzmann constant $k_{\rm B}$).  Assuming a pure Gaussian beam,
one finds a Rayleigh range $z_{\rm R}$ of 1.5\,mm. The axial and
radial oscillation frequencies in the harmonic approximation are
given by $\omega_{\rm z} = (2 U_0 / m z_{\rm R}^2)^{1/2}$ and
$\omega_{\rm r} = (4 U_0 / m w^2)^{1/2}$ with $m$ denoting the
mass of the atom.

\begin{figure*}
\center
\resizebox{0.75\textwidth}{!}{%
  \includegraphics{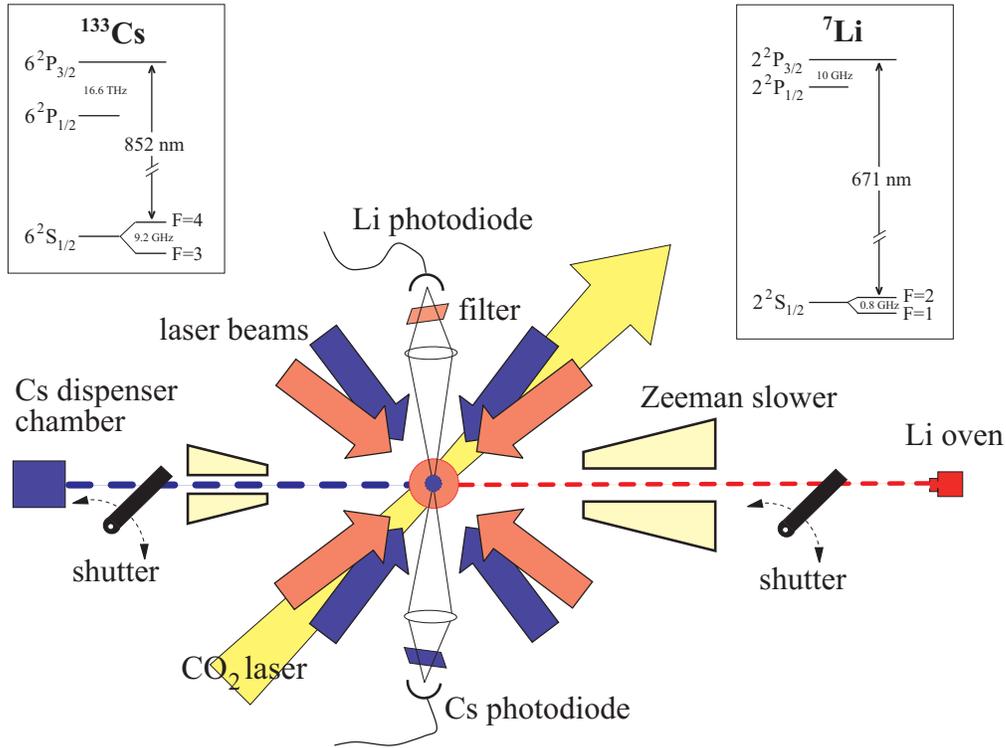}
}
\caption{Schematic setup of the experiment. The insets show the
relevant energy levels of lithium and cesium.}
\label{fig:setup}       
\end{figure*}

 Fig.~\ref{fig:setup} shows a schematic view of
the apparatus. Atoms are transferred into the dipole trap from
magneto-optical traps (MOT) for lithium and cesium which are
superimposed on the focus of the CO$_2$ laser beam. Both MOTs are
loaded from atomic beams that are cooled and decelerated by
decreasing-field Zeeman slowers which use the fringe fields of the
MOT magnetic quadrupole field for the final deceleration stage.
Fluorescence of the atoms in the MOT is detected by photodiodes
with appropriate filters to discriminate the resonance
fluorescence of Li at 671\,nm from the fluorescence of cesium at
852\,nm. The main vacuum chamber at a background pressure of about
$6 \times 10^{-11}$\,mbar is connected to the oven chambers by
tubes which are divided into differentially-pumped sections. Both
atomic beams can be interrupted by mechanical shutters.

\section{Storage in the quasi-electrostatic trap}
\label{sect:storage}
\subsection{Cesium}
\label{sect:Cs} To load cesium into the QUEST, a cloud of
typically $10^7$ atoms is first accumulated in the MOT with a
quadrupole field gradient of 26\,G/cm along the symmetry axis. The
atoms are further compressed and cooled by decreasing the detuning
of the MOT trapping laser from initially $-2\,\Gamma_\mathrm{Cs}$
to $-25 \Gamma_\mathrm{Cs}$ with respect to the
6$^2$S$_{1/2}$($F=4$) - 6$^2$P$_{3/2}$($F=5$) transition (natural
linewidth $\Gamma_\mathrm{Cs}/2 \pi = 5.3$\,MHz). After 5\,ms of
compression, the distribution of atoms in the MOT has a rms radius
of 400\,$\mu$m corresponding to an mean density of about
$10^{10}$\,atoms/cm$^3$. After the compression, the magnetic field
of the MOT is turned off, and the atoms are further cooled in a
blue optical molasses~\cite{hemmerich95} by detuning the laser
field to $+2 \Gamma_\mathrm{Cs}$ with respect to the
6$^2$S$_{1/2}$($F=4$) - 6$^2$P$_{3/2}$($F=4$) transition for
20\,ms.

The CO$_2$ laser beam is present during the whole loading phase.
Since the static polarizability of the 6$^2$P$_{3/2}$ state in
cesium is larger than the polarizability of the ground state by
around a factor of 4, the atomic resonance frequency of the cesium
D2 line is inhomogeneously shifted with a maximum value of roughly
$-10 \Gamma_\mathrm{Cs}$. However, the number of atoms in the MOT
is found to be independent of whether the CO$_2$ is turned on or
off.

\begin{figure}
\center
\resizebox{0.5\columnwidth}{!}{%
  \includegraphics{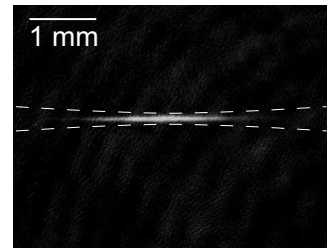}
} \caption{Absorption image of cesium atoms in the QUEST. The
dashed lines indicate the profile of the CO$_2$ laser beam}
\label{fig:abs}       
\end{figure}

After the cooling laser beams have been turned off, up to $10^6$
atoms remain trapped in the focus of the CO$_2$ laser beam at a
temperature of $30\,\mu$K and a peak density close to
$10^{12}$\,atoms/cm$^3$. Atoms are prepared in either the $F=3$ or
the $F=4$ hyperfine ground state by blocking the cooling laser
1\,ms after or before the repumping laser has been blocked,
respectively. To detect the optically trapped atoms, we either
recapture the atoms in the MOT and measure the intensity of the
fluorescence light, or we analyze the spatially resolved
absorption of a weak resonant probe beam passing through the
trapped atoms (absorption imaging~\cite{ketterle99}). The latter
method allows one to absolutely calibrate the atom number. In
Fig.~\ref{fig:abs}, the absorption image of the trapped Cs atoms
shows the expected elongated shape with an aspect ratio of
$\omega_{\rm ax}/\omega_{\rm rad} \simeq 44$. Note, that the
radial extension of the atomic cloud is not resolved by our camera
system.

\begin{figure}
(a)\\ \begin{center} \resizebox{0.7\columnwidth}{!}{%
  \includegraphics{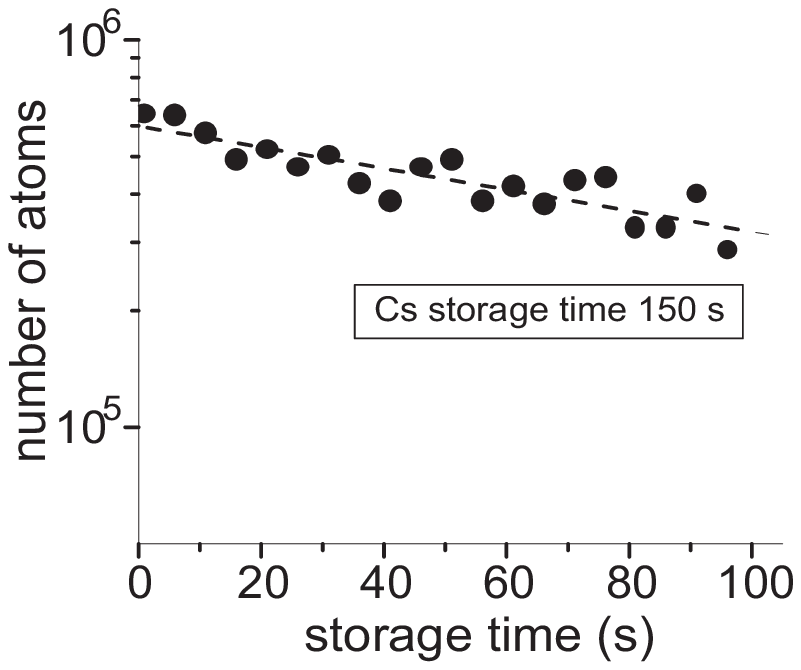}} \\ \end{center}

(b)\\
\begin{center} \resizebox{0.7\columnwidth}{!}{%
  \includegraphics{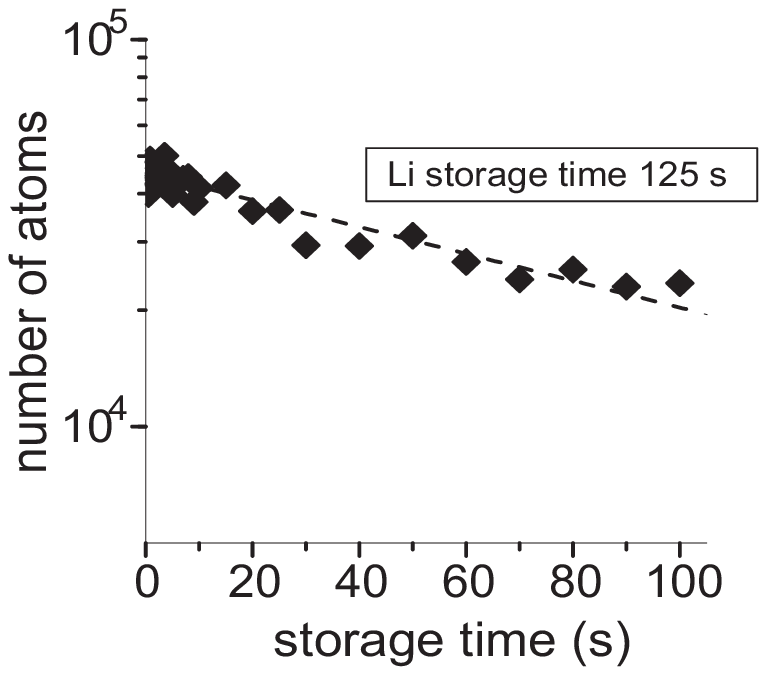}} \end{center}
\caption{Very long storage times for (a) cesium and (b) lithium in
the quasi-electrostatic trap. The atoms are optically pumped into
the lowest ground state.  The corresponding $1/e$ lifetimes are
150 $\pm 15$\,s for cesium and $125\pm 15$\,s for lithium.}
\label{fig:lifetime}       
\end{figure}

The evolution of the number of atoms in the dipole trap is
measured by recapturing the atoms back into the MOT after a
variable storage time. The number of atoms is determined from a
charge coupled device camera picture by integrating over the
fluorescence image, while subtracting a background image. We
expect the particle number determined by this method to be
accurate within a factor of two. Fluctuations in the loading flux
are cancelled by loading the MOT up to a certain density
(determined by the fluorescence signal) rather than for a fixed
period of time: A photodiode records the fluorescence of the atoms
in the MOT and triggers the transfer cycle as soon as it exceeds a
predefined discriminator level.
 The decay of the number of stored cesium atoms
is shown in Fig.~\ref{fig:lifetime}(a) from which one infers a
$1/e$ storage time of 150\,s. The very long storage time is purely
limited by collisions with hot atoms out of the background gas at
the base pressure of $6 \times 10^{-11}$\,mbar in our vacuum
chamber~\cite{engler00}. Laser-noise induced loss
mechanisms~\cite{savard97} can apparently be completely neglected.
Since Cs has a large scattering length the Cs gas thermalizes on a
time scale of $<0.1$ s. In contrast to previous experiments in a
shallower trap~\cite{engler00}, spontaneous evaporation of atoms
is exponentially suppressed by a factor $\exp(-\eta)$, where $\eta\equiv
\epsilon_t/k_BT$ is more than 20. Experimentally, we indeed
observe neither loss of Cs atoms nor reduction of the temperature
due to evaporation.

\begin{figure}
(a)\\ 
\begin{center} \resizebox{0.7\columnwidth}{!}{%
  \includegraphics{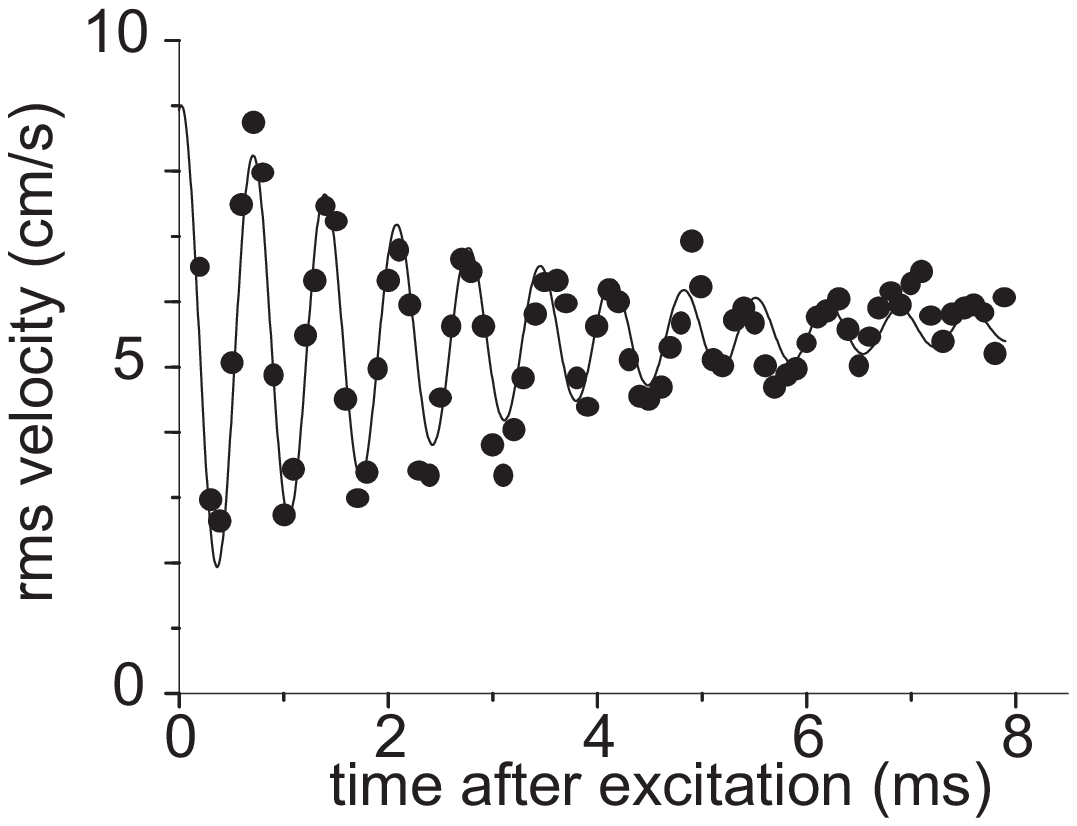}}\\ \end{center}
(b)\\ 
\begin{center} \resizebox{0.7\columnwidth}{!}{%
  \includegraphics{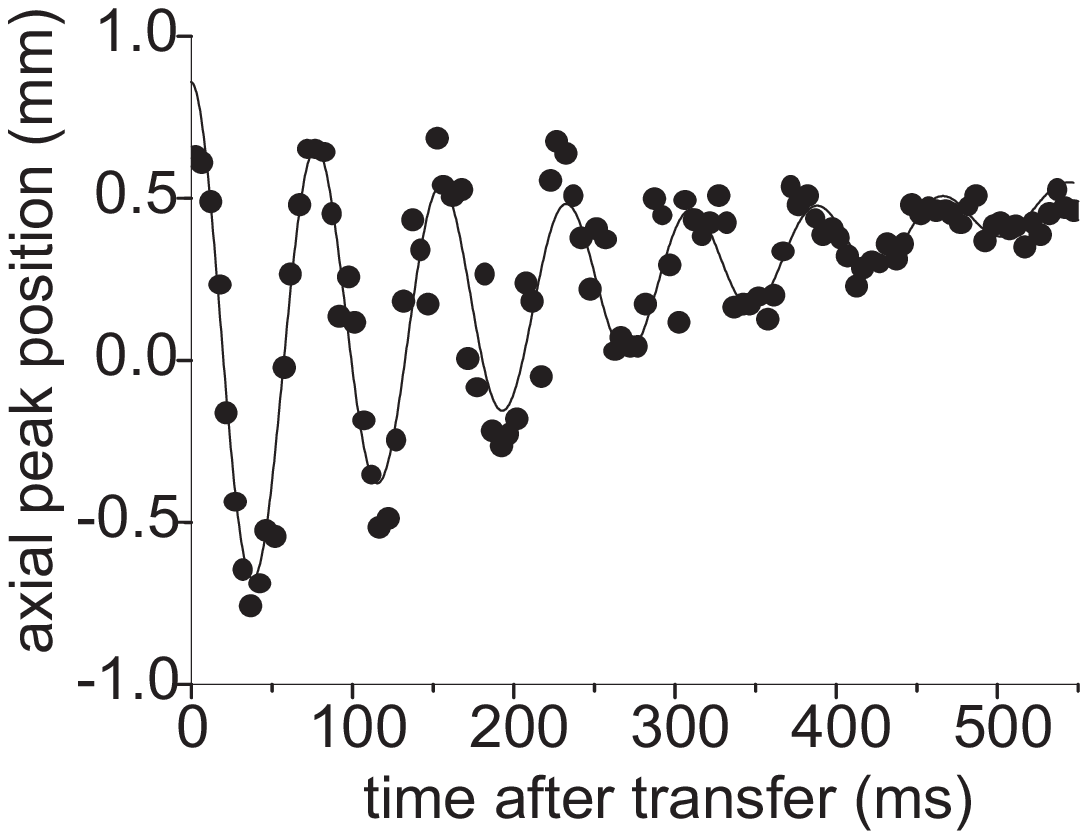}} \end{center}
\caption{Measurement of (a) radial, and  (b) axial oscillations of
Cs atoms in the quasi-electrostatic trap. The solid lines show the
fits to a damped sinusoidal oscillation. For the axial
oscillation, a constant drift of the focus position was included
into the fit.}
\label{fig:osc}       
\end{figure}

To determine the characteristic frequencies of the trap, we induce
axial and radial oscillations of the cloud of trapped atoms.
Radial oscillations are excited by a release-recapture cycle in
the optical trap. First, the laser is turned off (switch-off time
$<100$\,$\mu$s) which leads to a ballistic expansion of the atomic
cloud. After $\sim 0.5$\,ms, the cloud has expanded to about twice
its initial radial width. Then, the CO$_2$ laser is turned on
again and almost all atoms are recaptured by the dipole trap.
Through this process, the atoms have gained potential energy,
which is subsequently converted into kinetic energy. Thus, the
radial size of the atom cloud starts to oscillate at twice the
radial oscillation frequency. Since we can not fully resolve the
radial extension of the atom cloud with our absorption imaging
system, we measure the corresponding oscillation in the kinetic
energy. For this purpose, the CO$_2$ laser is turned off again
after a variable delay relative to the release-recapture cycle,
and the radial extension is determined through absorption imaging
after 1.2\,ms of ballistic expansion. Fig.~\ref{fig:osc}(a) shows
the oscillation of the root mean square velocity at a frequency of
1.6\,kHz. The corresponding oscillation frequency of the trapped
atoms of 0.8\,kHz is in reasonable agreement with the expected
radial oscillation frequency for a focused beam with a waist of
85\,$\mu$m.

Axial oscillations can easily be excited by loading at an axial
position off the focus. For this purpose, the relative position of
the MOT is shifted along the propagation direction of the
CO$_2$-laser beam. After transfer into the QUEST, the cesium cloud
starts to perform a damped oscillation at a frequency of 18\,Hz as
depicted in the upper graph of Fig.~\ref{fig:osc}. The damping of
the oscillation is caused by the anharmonicity of the confining
potential. Assuming a pure Gaussian beam, one would expect an
oscillation frequency about twice as large as the actual measured
value. We explain the discrepancy by aberrations in the optical
system.  Table~\ref{tab:parameters} summarizes the relevant trap
parameters for lithium and cesium.

\begin{table}
\label{tab:parameters}       
\begin{tabular}{llll}
\hline\noalign{\smallskip}  & trap depth & radial frequency
 & axial frequency
 \\
\noalign{\smallskip}\hline\noalign{\smallskip} Lithium &
390\,$\mu$K & 50\,Hz & 2.2\,kHz \\ Cesium & 1000\,$\mu$K & 18\,Hz
& 0.8\,kHz
\\ \noalign{\smallskip}\hline
\end{tabular}
\caption{Parameters of the quasi-electrostatic trap for lithium
and cesium. The Li frequencies are calculated from the measured Cs
frequencies.}
\end{table}

In the first few seconds after the CO$_2$ laser has been turned
on, we observe a small drift of the axial position of the laser
focus, which can be attributed to thermal effects in the laser
cavity and optics. To eliminate this drift of the focal position,
we leave the CO$_2$ laser permanently on while loading the MOT and
transferring the atoms.

\subsection{Lithium}

For cesium the thermal energy of the atoms in the MOT is much
smaller than the depth of the QUEST which results in a rather
large transfer efficiency. The situation is different for lithium.
Due to the large recoil of photons on lithium atoms, the thermal
energies in a lithium MOT are comparable or even larger than the
depth of the dipole trap resulting in a poor phase-space overlap
and thus a small transfer efficiency.

The lithium MOT is also loaded from a Zeeman-slowed atomic beam.
Details of the MOT system are described in Ref.~\cite{schune98}.
The largest numbers of Li atoms are accumulated at relatively
large detunings of the MOT lasers acting on the two hyperfine
components of the Li $D2$ line. After loading typically $10^7$
atoms into the MOT, the atomic cloud is compressed for 4\,ms by
switching the detunings closer to resonance
($-1\,\Gamma_\mathrm{Li}$ with $\Gamma_\mathrm{Li}/2\pi =
5.9$\,MHz). Due to the unresolved hyperfine structure of the
excited state, sub-Doppler cooling mechanisms are inefficient for
lithium and temperature is essentially determined by the limit for
Doppler cooling (140\,$\mu$K at low saturation). To attain a
temperature close to that limit,
 the MOT magnetic
field is turned off and the MOT lasers are attenuated
to 20\% of their full power which leads to a final temperature of
350\,$\mu$K and a cloud size of 800\,$\mu$m after typically 1\,ms.
In contrast to cesium, the light shift of the CO$_2$ laser beam on
the $D2$ resonance line is negligible since the static
polarizability of the ground and excited state differ by only
25\%.

After all MOT laser beams have been shut off, about 0.5\% of the
Li atoms are captured by the QUEST with a peak density of
$\sim 10^{10}$\,atoms/cm$^3$. Thermalization of these atoms is extremely
slow: estimated collision times based on the s-wave scattering
lengths for Li are exceeding 100 seconds. On the time scales of
the experiments discussed in the following, lithium can therefore
not be assumed to be in thermal equilibrium.

Due to the small optical density of the sample, we were so far not
able to directly measure the distribution of lithium atoms in the
trap by absorption imaging techniques. Instead, we have performed a
Monte-Carlo simulation of the transfer process which provides us
with the spatial and energy distributions. The energy distribution is highly
non-thermal with most atoms populating high lying energy states in
the potential because of the large initial thermal energy relative
to the trap depth. The small rate for elastic collisions inhibits
evaporation of these loosely bound atoms. The measurement of the
number of trapped lithium atoms as a function of storage time
depicted in Fig.~\ref{fig:lifetime} thus shows a pure exponential decay
with a time constant of 125\,s with no indication for evaporation
of particles from the trap. On the basis of the above estimate for
the thermalization times, we assume the energy distribution hardly
changes with time.

\subsection{Combined transfer}

To load a mixture of lithium and cesium, one could start from a
two-species MOT~\cite{schloeder99}, but in this case the numbers
of atoms and the densities are severely limited by light-assisted
inelastic collisions between the two species. However, our
measurements with a 2-species MOT showed that collisions involving
optically excited Li (Li$^*$) and ground-state Cs are strongly
suppressed by optical shielding through the repulsive
van-der-Waals interaction in the Li$^*$-Cs
manifold~\cite{schloeder99}. It is therefore favorable to first
optically cool and transfer Cs into the QUEST, and then load the
Li MOT. A typical loading time of the Li MOT is 4\,s. In order to
exclude any loss processes by inelastic collisions, we spatially
separate the Li MOT from the optically trapped Cs sample with a
small bias magnetic field (3\,G) shifting the centre of the Li MOT
by about 2\,mm. In this way, we observe no loss of Li and only a
50 \% loss of Cs in the combined transfer, as compared to the
separate trapping of only one species. A detailed investigation of
the combined transfer will be published elsewhere.

\section{Spin-changing lithium-cesium collisions}
\label{sect:spinchange} Two atoms approaching each other in their
electronic ground state may either collide elastically with the
total energy being conserved, as will be discussed in the next
Section, or they may undergo an exoergic collision transforming
internal energy into kinetic energy of the collision partners. The
latter process leads to trap loss if the kinetic energy release is
larger than the escape energy of the trap. Optical dipole traps
are perfectly suited to study inelastic processes among trapped
atoms since the optical trapping force is conservative with a
well-defined potential, the investigation may be performed in zero
magnetic field or in a controlled homogeneous field, and the atoms
can be prepared in any internal state by appropriate optical
pumping methods. Up to now, no experiment on ultracold inelastic
collisions between two different atomic species in the electronic
ground state has reported.

\begin{figure}
\center
\resizebox{0.7\columnwidth}{!}{%
  \includegraphics{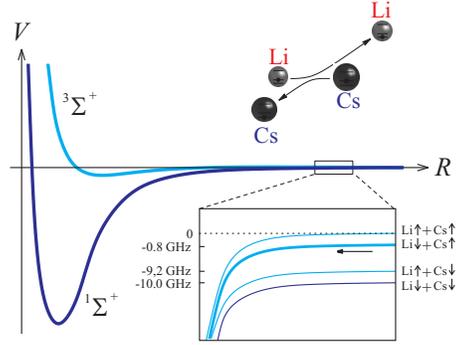}
} \caption{Ground-state potential energy curves  for the heteronuclear dimer LiCs.  The
  inset shows an enlargement at long range.  The separated atom
  hyperfine levels are characterized by the total angular momentum $F$
  ($F_{{\rm Cs}\downarrow}$= 3 and $F_{{\rm Cs}\uparrow}$= 4 for Cs,
  $F_{{\rm Li}\downarrow}$= 1 and $F_{{\rm Li}\uparrow}$= 2 for Li). The
  arrow indicates the incoming channel.}
\label{fig:lics_hype_potential}
\end{figure}

For ground-state Li and Cs atoms in an optical dipole trap, the
essential inelastic processes are collisions that change the
hyperfine quantum numbers $F$ and $m_F$. The spin-changing
collision process between lithium and cesium can be written
symbolically as
\begin{eqnarray*}
{\rm Li}(F\!=\!1) + {\rm Cs}(F\!=\!4) & \rightarrow & {\rm
LiCs}(^1{\rm \Sigma}_g \ {\rm or} \  ^3{\rm \Sigma}_u)\\ & \rightarrow & {\rm
Li}(F\!=\!1) + {\rm Cs}(F\!=\!3) + \Delta \epsilon_{\rm kin} \, .
\end{eqnarray*}
The potential curves for the singlet and triplet ground states of
the Li-Cs pair are schematically depicted in
Fig.~\ref{fig:lics_hype_potential}. Spin-changing collisions
exiting at a lower energetic asymptote than the entrance channel
lead to a transformation of the internal energy into kinetic
energy $\Delta \epsilon_{\rm kin}$ of the collision partners, and
consequently to loss from the trap. Although lithium takes the
major share of the kinetic energy gain, both atoms escape from the
QUEST due to the small trap depth as compared to the hyperfine
energy.

Inelastic collisions of trapped ground-state atoms have been extensively
studied theoretically and experimentally since the early days of
magnetic traps~\cite{silvera86}. Hyperfine-changing collisions
represent the major limitation on achievable densities and/or
storage times in magnetic traps. These collisions in magnetic
traps are almost unavoidable, since atoms can not be trapped in
the lowest energetic state. In particular, spin-changing
collisions impede sympathetic cooling between different atomic
species in magnetic traps.

Exoergic ground-state collisions also turned out to be an
important loss mechanism in shallow light-pressure traps.  They
were observed and analyzed for homonuclear
collisions~\cite{weiner99}, as well as for mixtures of different
elements~\cite{Bagnato,Bigelow,schloeder99}. However, as has been
pointed out recently~\cite{gensemer00}, the presence of
near-resonant light strongly modifies the dynamics of the
collision process which can lead to a dramatic change in the
measured rate coefficients. Therefore, care should be taken to
interpret trap loss and the corresponding rate coefficients in
shallow light-pressure traps as an indication of pure ground-state
collisions. In a purely conservative trap such as the QUEST the
situation is much better defined since the trapping light has
negligible influence on the collision process.

\begin{figure}
(a)\\
\begin{center} \resizebox{0.8\columnwidth}{!}{%
  \includegraphics{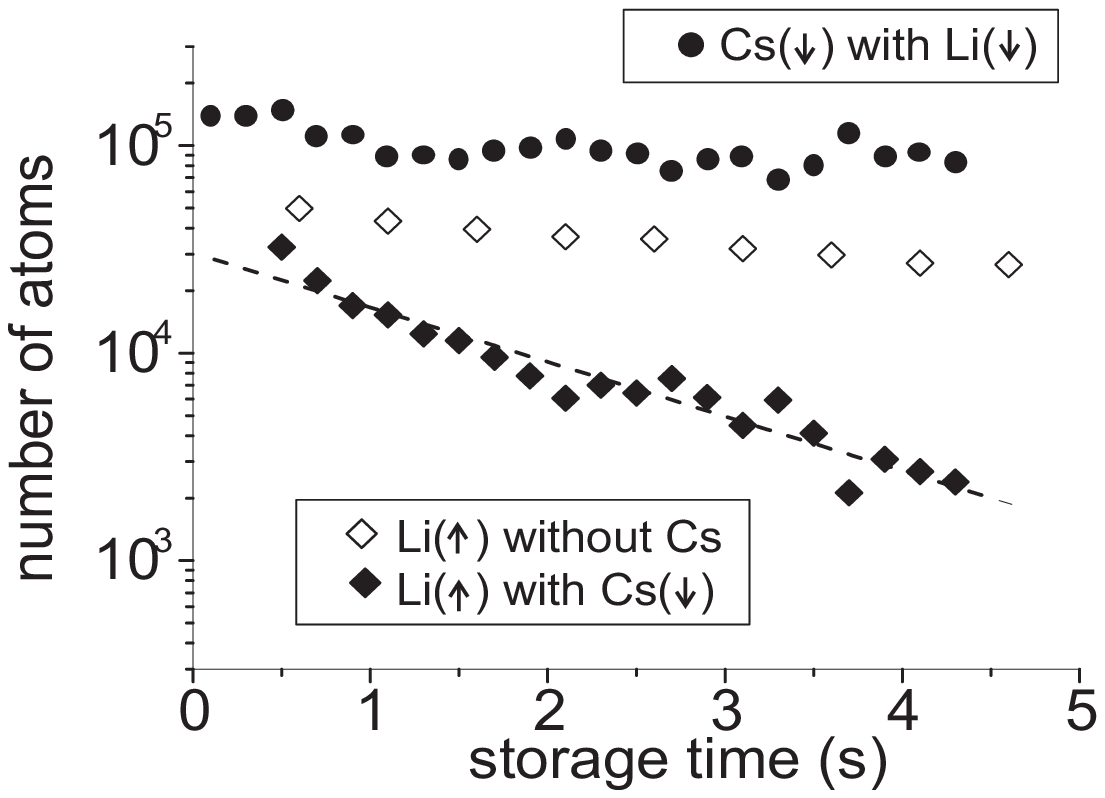}}\\ \end{center}
(b)\\
\begin{center} \resizebox{0.8\columnwidth}{!}{%
  \includegraphics{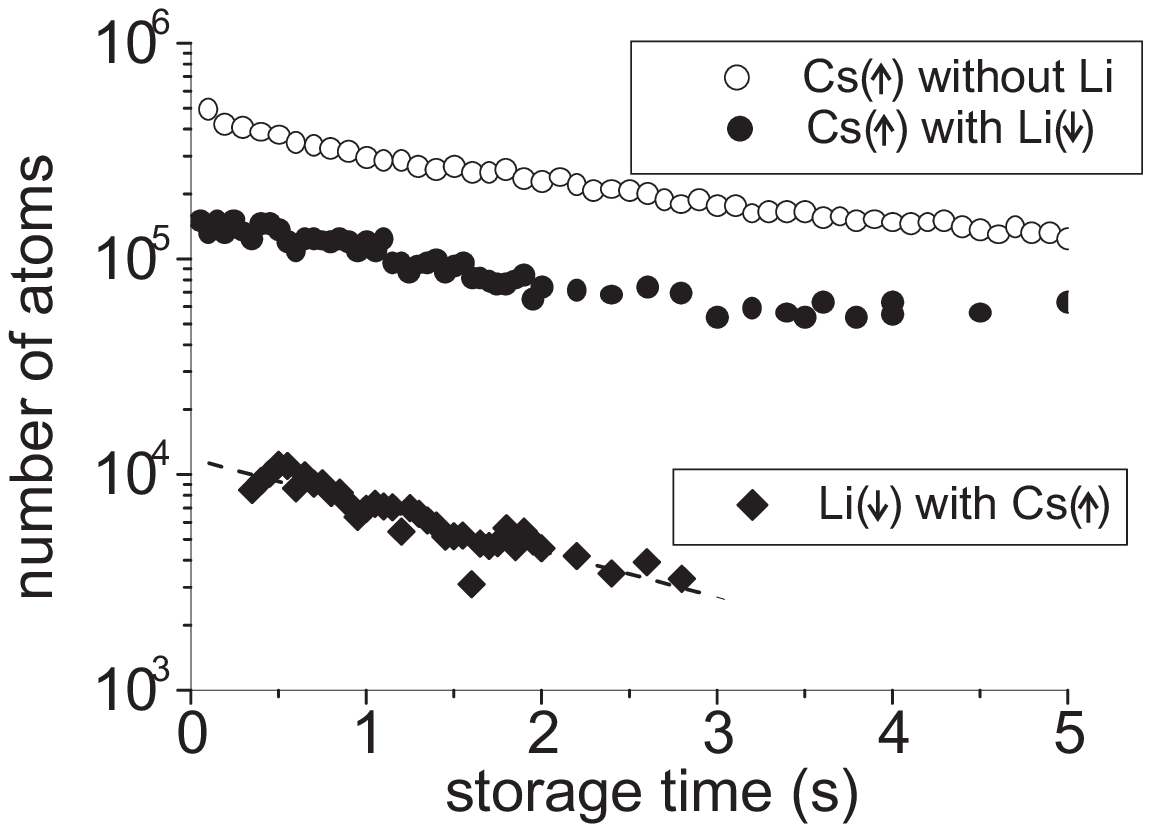}} \end{center}
\caption{Exoergic spin-changing collisions between lithium and
cesium resulting in trap loss. (a): Li $2^2S_{1/2}(F_{{\rm
Li}\uparrow}$=2) and Cs $6^2S_{1/2}(F_{{\rm Cs}\downarrow}$=3).
(b): Li $2^2S_{1/2}(F_{{\rm Li}\downarrow}$=1) and Cs
$6^2S_{1/2}(F_{{\rm Cs}\uparrow}$=4). Here, $\uparrow$ stand for
the upper hyperfine ground state and $\downarrow$ stands for the
lowest-energy hyperfine state. The dashed lines indicate the fit
to an exponential decay yielding a time constant of 1.7\,s and
1.9\,s for the upper and lower graph, respectively.}
\label{fig:HeteroHypChange}       
\end{figure}

To qualitatively investigate whether hyperfine-changing collisions
between Li and Cs can be observed with the particle numbers and
densities available in our QUEST, we have pumped one of the
species into the upper hyperfine ground state and recorded the
effect on the storage time of the other species, which was
prepared in the lower hyperfine state.  The result of these
experiments is depicted in Fig.~\ref{fig:HeteroHypChange}. From
the timescale of the graph it immediately becomes apparent, that
the presence of a hyperfine-excited atomic gas reduces the storage
time of the species in the lowest energetic state by more than an
order of magnitude as compared to the rest-gas limited lifetimes
(see Fig.~\ref{fig:lifetime}).

In Fig.~\ref{fig:HeteroHypChange}(a), we show data for a
combination of
 Li in the upper hyperfine state [Li $2^2S_{1/2}(F_{{\rm Li}\downarrow}$=2)], with
 Cs in the lower hyperfine ground state [Cs $6^2S_{1/2}(F_{{\rm
Cs}\uparrow}$= 3)]. Without the presence of Cs, the Li slowly
decays through exoergic Li-Li collisions. The effect of collision
with restgas atoms can be neglected on this timescale. In presence
of Cs, the Li decay is much faster. Spin-exchange between Li and
Cs atoms is the dominant process for these interspecies
collisions. An enhanced Li-Li inelastic collision rate due to an
increased Li density after thermalization with the Cs may also
play a minor role, but this would lead to a non-exponential decay
of the Li, while we observe an exponential decay with a time
constant of 1.7 s. The effect of spin-changing Li-Cs collisions on
the Cs is not observable as the Cs atoms outnumber the Li atoms.
As the lithium distribution is not known, and may even be
nonconstant due to elastic collisions happening simultaneously, we
can currently only estimate the order of magnitude of the
inelastic rate constant $\beta_{\rm Cs\downarrow Li\uparrow} \sim
10^{-10}$~cm$^3/$s. We are currently implementing an improved
absorption imaging system for Li in our apparatus, which will
allow us to measure the density distribution of the Li atoms und
thus allow for quantitative measurements of the rate coefficients.

In Fig.~\ref{fig:HeteroHypChange}(b), we investigate the influence
of Cs in the upper hyperfine ground state [Cs $6^2S_{1/2}(F_{{\rm
Cs}\uparrow}$=4)] on Li in the lower hyperfine state [Li
$2^2S_{1/2}(F_{{\rm Li}\downarrow}$=1)]. Cs itself decays due to
exoergic Cs-Cs collisions. The Li, however, is lost on a much
shorter timescale of 2\,s.  Due to the unknown Li distribution, we
can only offer an estimate of the rate constant of the exoergic
collisions of $\beta_{\rm Cs\uparrow Li\downarrow}=\sim
10^{-10}$~cm$^3/$s. As one might expect, both rate constants are
of the same order of magnitude.

\section{Sympathetic cooling with slow sympathetic evaporation}
\label{sect:symp}

In case both atomic gases are in the lower hyperfine states [Li
$2^2S_{1/2}(F_{{\rm Li}\downarrow}$=1) and Cs $6^2S_{1/2}(F_{{\rm
Cs}\downarrow}$=3)], only elastic collisions are possible. These
collisions lead to energy transfer between the distributions,
which in turn leads to two different phenomena: The most obvious
one is the equalization of the temperatures of the two gases,
which is referred to as sympathetic cooling. A second phenomenon
is the evaporative loss of one or both species through collisions
with the other species, which we call sympathetic evaporation.


\subsection{Sympathetic cooling}

For the case where all relevant temperatures are well below the
trap depth, the dynamics are relatively simple: No atoms evaporate
and the temperatures of the two components approach each other. A
quantum-mechanical treatment of sympathetic cooling has been given
in Refs.~\cite{lewenstein95,papenbrock01}. For the Boltzmann
regime, an analytical formula for the evolution of the
temperatures in a mixture of equal mass atoms was derived in
Ref.~\cite{AspectPRA2001}. In the same paper, the approach is
extended towards atoms with different mass, however, the
dependence of the cooling efficiency on the mass ratio is derived
under the assumption of equal trap frequencies. This assumption
corresponds to an implicit mass dependence of the trap potential,
which is not present for realistic traps, and which leads to a
strongly overestimated reduction of the cooling efficiency at
large mass ratio. Below, we adapt the approach of
Ref.~\cite{AspectPRA2001} by assuming an explicit trap potential,
so that our results remain valid for large mass ratios. The
average energy transfer per collision can be calculated from
simple kinematics,
\begin{equation}
\Delta E_{1 \to 2} = \xi  k_B (T_1-T_2); \;\; \xi\equiv \frac{4 m_1
m_2}{(m_1+m_2)^2}.
\end{equation}
The energy transfer is reduced by a factor $\xi$
with respect to collisions between atoms with equal mass.
For equal mass atoms in a harmonic trap, approximately 3 collisions are needed for
thermalization (see \cite{AspectPRA2001} and references therein),
as the heat capacity per atom is $3 k_B$.
For $^{133}$Cs and $^7$Li, we need $3/\xi \approx 15$ collisions
to thermalize.

The  Cs-Li collision rate depends on the spatial overlap between
the thermal distributions,
\begin{equation}
\Gamma_{coll}= \sigma_{12} \bar{v} \int d^3 x\, n_1(\mathbf{x}) n_2(\mathbf{x})
\end{equation}
Here $\bar{v}$ is the mean thermal relative velocity,
\begin{equation}
\bar{v}=\left[
\frac{8 k_B}{\pi}(\frac{T_1}{m_1}+\frac{T_2}{m_2})
\right]^{1/2}
\end{equation}
We assume a harmonic trapping potential,
$V_i(\mathbf{x})=(\mbox{\boldmath{$\alpha$}}_i\cdot\mathbf{x})^2$,
where $i=1,2$,
and  {\boldmath{$\alpha$}$_i$} is a species-dependent vector of
force constants. We will assume the trap constants are linearly
dependent, i.e.
{\boldmath{$\alpha$}$_2$}$=$$\beta${\boldmath{$\alpha$}$_1$}$=$$\beta${\boldmath{$\alpha$}}.
This allows us to write the collision rate as
\begin{equation}
\label{eq:gammacolln}
\Gamma_{coll}= \frac{\sigma_{12} \bar{v} N_1 N_2}{
\pi^{3/2}k_B^{3/2}} \alpha_x\alpha_y\alpha_z
\left(
T_1+T_2 \beta^{-2}
\right)^{-3/2}.
\end{equation}

The timescale of thermalization can be obtained from the
differential equation for $\Delta T=T_2-T_1$:
\begin{eqnarray}
\label{temperatureevolution}
\tau^{-1}&=&\frac{d(\Delta T)}{\Delta T d t}=
\frac{4 (N_1+N_2)\sigma_{12}\xi \alpha_x\alpha_y\alpha_z}{3
\pi^{2}k_B^{}}\times
\\
&& \nonumber
\frac{(T_1/2m_1+T_2/2m_2)^{1/2}}{(
T_1+T_2 \beta^{-2}
)^{3/2}}
\end{eqnarray}
For equal masses and trap frequencies we reproduce the result of
Ref.~\cite{AspectPRA2001}.

For the initial conditions of our experiment
we obtain an approximate initial cooling
time constant, where we neglect the influence of the Cs temperature since Cs is
much colder, heavier and more tightly confined than the Li.
\begin{eqnarray}
\label{initial}
\tau_{\rm init}^{-1}&=&
\frac{4 (N_{\rm Li}+N_{\rm Cs})\sigma_{\rm LiCs}\xi \alpha_x^{\rm Li}\alpha_y^{\rm Li}
\alpha_z^{\rm Li}}{3
\pi^{2}k_B T_{\rm Li}(2m_{\rm Li})^{1/2}}.
\end{eqnarray}
In the final stage of sympathetic cooling the final temperature
$\bar{T}=(N_1T_1+N_2T_2)/(N_1+N_2)$ is approached exponentially, with a time constant
\begin{eqnarray}
\label{tfinal}
\tau_f^{-1}&=&
\frac{4 \sqrt{2 \xi} (N_1+N_2)\sigma_{12}\alpha_x\alpha_y\alpha_z}{3
\pi^{2}k_B^{}\bar{T} (m_1+m_2)^{1/2}}
(
1+ \beta^{-2}
)^{-3/2}.
\end{eqnarray}
The effect of the mass difference on the
thermalization time  is a factor $\xi^{-1/2}\sim 2.3$ in the time
constant.  At large mass difference, the thermalization efficiency
\emph{per collision} is reduced by a factor $\xi$, but the collision
frequency depends on $\bar{v}$, which is proportional to $\xi^{-1/2}$.

\subsection{Sympathetic evaporation}
Under the circumstances in our experiment, we expect evaporation
of the Li atoms especially during the first few collision times,
as the initial Li energy distribution has a width comparable to
the trap depth. Once the gases are thermalized, Li evaporation is
exponentially suppressed, but still may occur due to the high
Li-Cs collision rate. Cs evaporation is always strongly suppressed
as the Cs trap depth is much larger than the relevant
temperatures. We can approximately describe the regime of slow
evaporation by assuming a truncated Boltzmann distribution
\cite{Luiten} for the Li atoms. The number of evaporation events
per second is the number of collisions multiplied by the
probability $P_{\rm evap}$ that a collision leads to evaporation.
This probability can be found from a detailed-balance approach in the following way:
If the gas is in full thermal equilibrium (i.e. at the same temperature and chemical potential)
with a gas of identical untrapped
atoms, evaporation will be exactly compensated by "atom capture"
events.
The untrapped atoms will have a density $n_u({\bf r})$ which is to good approximation
$n({\bf r}) \eta^{1/2} \exp(-\eta)$. In case $\eta\xi \gg 1$,
practically every collision of a trapped with an untrapped atom
will lead to capture of the untrapped atom, i.e., the number of
capture events is well approximated as the number of collisions between untrapped
atoms and trapped atoms, which is a fraction $\eta  \exp(-\eta)$
of the total number of collisions. It follows from the detailed
balance approach that this equals the number of evaporation
events, so that we find the following approximation:
\begin{equation}
\label{eq:Pevap}
 P_{\rm evap} \sim \eta \exp(-\eta).
\end{equation}
This evaporation probability leads to an evaporation rate of $\Gamma_{\rm evap} = \Gamma_{\rm col} P_{\rm
evap}$. The amount of energy lost in an evaporation event is of
order $(\eta +1) k_B T$, while the Cs atom
that is involved in the last collision only loses a small
amount of energy, of order $k_B T$. The Li distribution on the other
hand loses a particle and an energy of $\sim \eta k_B T$, leading
to a decrease of the mean energy per Li atom of
\begin{equation}
\frac{\Delta (e_{\rm Li})}{k_B T}=\frac{(\eta - 3/2 -
\delta)}{N_{\rm Li}}.
\end{equation}
Here
$\delta$ represents the average potential energy, $\delta=0$
for a box potential and $\delta=3/2$ for a harmonic trap.
Since the evaporation cools the Li more than the Cs,
we expect that evaporation leads to a small difference in mean
energy
between the two components, of order
\begin{equation}
\label{evapdeltat}
\frac{e_{\rm Cs}-e_{\rm Li}}{\bar{e}} \approx \frac{\eta}{2 \xi} e^{-\eta} \left(
\eta-3/2-\delta- \eta \xi
\right).
\end{equation}
Here $\bar{e}=(N_{\rm Cs} e_{\rm Cs}+N_{\rm Li} e_{\rm Li})/(N_{\rm Cs}+N_{\rm
Li})$.
The evaporation-induced energy difference for a mixture of equal mass atoms is of
the same order of magnitude as the difference in mean energy
between a truncated Boltzmann distribution and the full thermal
distribution described by the same temperature \cite{Luiten}. For
atoms with different masses, the thermalization is slower by a
factor $\xi$, whereas the evaporation rate is practically
independent of the mass difference. This leads to an enhancement
of the
temperature difference in evaporating mixtures, as we have
observed in simulated evaporation experiments.

\begin{figure}
\center \resizebox{\columnwidth}{!}{\includegraphics{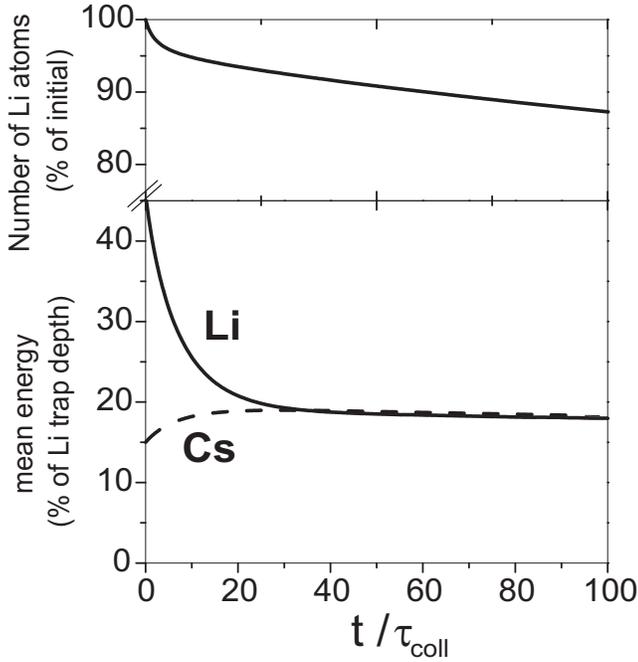}
} \caption{Evolution of the particle number and internal energy in
a simulated
 thermalization experiment. The number of Cs atoms is  constant at
5 times the initial number of Li atoms. The mean energies
  cross over after 45 collisions, see text.}
\label{fig:SimulEvo}       
\end{figure}

\begin{figure}
\center
\resizebox{\columnwidth}{!}{\includegraphics{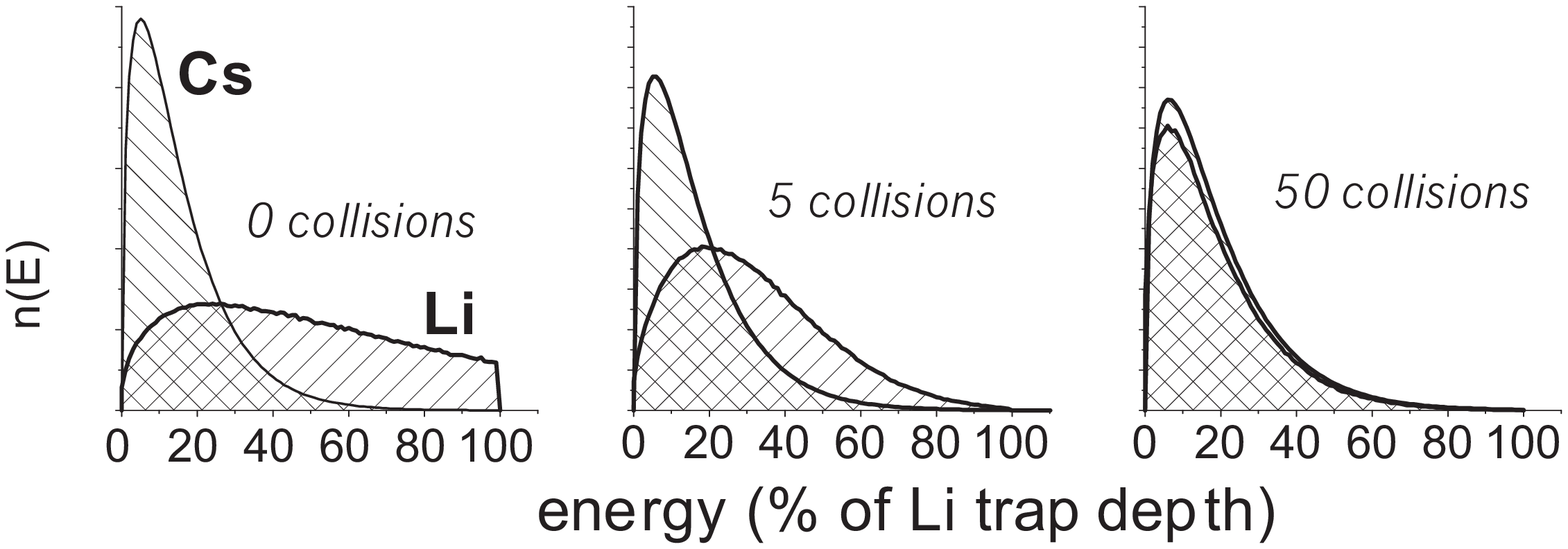}
} \caption{Simulated evolution of the Li and Cs distributions. The
left pane shows the initial Li distribution, which is a strongly
truncated Maxwell-Boltzmann distribution. The Cs distribution is
shown reduced in height by a factor 5.}
\label{fig:SimulDis}       
\end{figure}
\subsection{Numerical simulation}
The initial phase of thermalization, in which the distributions
are strongly non-thermal, is difficult to describe in an
analytical model.
To obtain understanding of
all phases of the thermalization process we made
a numerical simulation of Li-Cs thermalization in
a finite depth ``box'' potential well.
We assume that due to the large scattering length the Cs distribution
remains thermal at all times, while we numerically integrate the
Boltzmann equation for the Li distribution.
 In Fig.~\ref{fig:SimulEvo} we show the evolution of the mean energies
 and particle
numbers, where a broad initial distribution for the Li atoms was
chosen.
The evolution of the
distributions is shown in Fig.~\ref{fig:SimulDis}.

In the initial phase of the thermalization process, the first $\sim 5$ collisions
in Fig.~\ref{fig:SimulEvo},  the Li has an
internal energy per particle comparable to the trap depth, so that
evaporation is very fast.
After this rapid evaporation, a thermalization phase sets in (5 to
50 collisions) in which the mean energies rapidly converge.
Finally, after 50 collisions a quasi-stationary
phase with slow sympathetic evaporation and only slowly varying,
almost equal temperatures sets in.
We observe that although we start with $T_{\rm Li}>T_{\rm Cs}$,
after $\sim 50$ collisions the evaporating Li gas has a lower average
energy per particle than the Cs, reasonably well described by Eq.(\ref{evapdeltat}).
%
\begin{figure}
\center
\resizebox{0.75\columnwidth}{!}{\includegraphics*{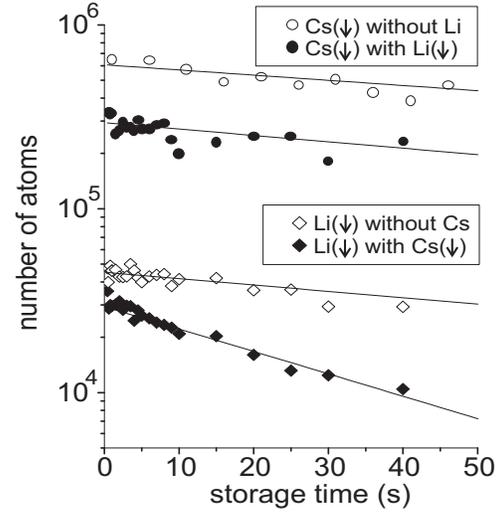}}
\caption{Evolution of the number of trapped Li (F=1) and Cs (F=3)
atoms in simultaneous trapping experiments (closed symbols). For
comparison, the open symbols show the evolution of separately
trapped gases. Thin lines: Exponential fits. Time constants are
150 s for Cs with or without Li, 125 s for Li without Cs, and 35 s
for Li trapped simultaneously with Cs.}
\label{fig:MeasEvo}       
\end{figure}

\subsection{Comparison with measurements}
A complete characterization of the thermalization process requires
measurement of all relevant thermodynamic variables: the Li and Cs
temperatures and particle numbers. Unfortunately, presently the Li
temperature cannot be measured, and as $N_{\rm Cs}\gg N_{\rm Li}$,
the effect of sympathetic cooling on the Cs temperature is too
small to be measured accurately.

The evolution of the number of Li atoms in the trap is shown in
Fig.~\ref{fig:MeasEvo}. When stored separately, both Li and Cs
gases decay due to rest gas collisions in $125$ and 150 s
respectively. There is no indication of evaporation in either
case: Cs does not evaporate as its temperature is far below the
trap depth. Li evaporation is energetically possible but the cross
section for Li-Li collisions is so small that the characteristic
timescale is $> 1000$ s.

When the gases are stored simultaneously, the Li atom number
shows a non-exponential initial behavior followed by an approximately
exponential decay with a characteristic time of 35 s, which is much faster
than the 125 s rest gas induced decay. We interpret the initial
loss as due to evaporation in the first few Li-Cs collisions (cf. Fig.~\ref{fig:SimulEvo}),
and the slow
decay as due to sympathetic evaporation after thermalization. Equation~(\ref{eq:Pevap})
now allows us to estimate the collision frequency. From the calculated Li trap depth and
the measured Cs temperature after 20 s, $T_{\rm Cs}=44(5)$ $\mu$K we find
$P_{\rm evap}\sim 1.5 \times 10^{-3}$. This leads to an estimated
collision rate per Li atom of $\sim 20$ s$^{-1}$, and finally through
Eq.(\ref{eq:gammacolln}) we find the estimated cross section
\begin{equation}
\sigma_{\rm CsLi}\sim 5 \times 10^{-12} {\rm cm}^{2}.
\end{equation}
From the collision rate we infer a
cross-species thermalization time of approximately 1 s, which
is much shorter than the rest gas limited storage times in our
trap.
Therefore, we conclude that
sympathetic cooling of Li by Cs is a viable method to produce ultracold
Li gas.

\section{Conclusion}
In conclusion, we have simultaneously stored ultracold lithium and
cesium atoms in a conservative optical dipole trap. Exoergic
spin-changing collisions lead to trap loss on a time scale of a
few seconds at densities of $10^{10}$ atoms/cm$^3$ and $10^{12}$
atoms/cm$^3$ for lithium and cesium, respectively. Under
conditions where exoergic collisions are impossible, a slow loss
of Li atoms is observed, which we attribute to sympathetic
evaporation through elastic Li-Cs collisions. The interpretation
in terms of a model for sympathetic cooling and evaporation
provides an estimated cross-thermalization time of one second.

The interpretation of the observations in terms of sympathetic
cooling should be confirmed by measurements of the temperature
evolution and phase space density. Such measurements are currently
in preparation. Temperature data will also enable us to measure
the elastic cross section $\sigma_{CsLi}$ for cold, elastic Li-Cs
collisions more accurately. Optimization of the Cs cooling process
is certainly possible and will lead to lower final temperatures of
the Li gas.

Repeated pulsed optical cooling of the Cs may be useful to remove
thermal energy from the system without loss of atoms. This method
offers a possible route towards quantum degeneracy of Li without
evaporation. The de~Broglie wavelength of Li is larger than that
of Cs at the same temperature by a factor of $4.4$, which makes it
possible to achieve quantum degeneracy with either the fermionic
or the bosonic Li isotopes in thermal contact with optically
cooled Cs.  For $10^6$ Li atoms in our trap we find a degeneracy
temperature
 $T_\mathrm{deg} \simeq \hbar (\omega_x\omega_y \omega_z \, N_\mathrm{Li})^{1/3}/k_B = 3
 \, \mu$K, well within the range of standard optical cooling
 techniques for Cs. But even at the temperatures we reach presently, sympathetic cooling
significantly increases the Li density in the centre of the trap.
This is an essential step towards creation of cold heteronuclear
molecules, as all proposed formation mechanisms, in particular
photoassociation, rely on a high pair densities of the two
species.

\section*{Acknowledgements}
This work has been supported in part by the Deutsche
Forschungsgemeinschaft. A.M. is supported by a Marie-Curie
fellowship from the European Community programme IHP under
contract number CT-1999-00316. We are indebted to D.~Schwalm for
encouragement and support. We thankfully acknowledge many fruitful
discussions with  A.N.~Salgueiro and H.A. Weidenm\"uller.

%

\end{document}